\documentstyle[twoside,fleqn,espcrc2]{article}
\def\NP#1#2{ Nucl.Phys. B#1 (#2)}
\def\PL#1#2{ Phys.Lett. B#1 (#2)}

\newcommand{\2}{{\hat sl}_2}

\newcommand{\bx}{\bar x}
\newcommand{\bz}{\bar z}

\def\fzx#1{\Phi^{j_{#1}}(x_{#1},\bx_{#1},z_{#1},\bz_{#1})}

\newcommand{\AmS}{{\protect\the\textfont2
  A\kern-.1667em\lower.5ex\hbox{M}\kern-.125emS}}
\title{Notes on SL(2) conformal fields theories. \\ Exact solution and 
applications
\thanks{Talk presented at the International
Symposium on the Theory of Elementary Particles, Buckow 1996}.}
\author{Oleg Andreev\thanks{On leave from Landau Institute for 
Theoretical Physics, Moscow}\thanks{e-mail address: 
Oleg.Andreev@peterpan.ens.fr} \\
Laboratoire de Physique Th\'eorique de l'\'Ecole Normale Sup\'erieure
\thanks{Unit\'e Propre du Centre National de la Recherche Scientifique,
associ\'ee \`a l'\'Ecole Normale Sup\'erieure et \`a l'Universit\'e 
de Paris-Sud.} ,\\
24 rue Lhomond, 75231 Paris C\'EDEX 05, France}

\begin{document}

\begin{abstract}
In these notes I briefly outline $SL(2)$ degenerate conformal field
theories and their application to some related models, namely 2d
gravity and $N=2$ discrete superconformal series.
\end{abstract}

\maketitle

\section{Introduction}

Since the seminal work of Belavin, Polyakov and Zamolodchikov
\cite{BPZ}, where a general approach to two-dimensional conformal
field theories was proposed, there has been much progress in understanding
these theories. However the full solutions were found only for
relatively few theories. The most famous examples are the diagonal
minimal models and $SU(2)$ WZW models \cite{DF,FZ}. One motivation for
my research was to extend this set by solving $SL(2)$ degenerate
conformal field theories. These theories contain, as a subclass,
$SU(2)$ models. Another motivation was to try to get information on
more complicated models using a progress with $SL(2)$ ones.

The outline of these notes is as follows.
\newline In section two I give a more formal discussion of the basic
points relevant for $SL(2)$ degenerate conformal field theories. Next,
in sections three and four, I present explicit examples of application
of the results described in section two to 2d gravity and to some $N=2$
discrete superconformal series. Finally, in section five I offer my
conclusions and mention a few important problems. 

\section{SL(2) degenerate conformal field theories}

The theories have $\2\oplus\2$ algebra as the symmetry algebra. The
commutation relations for the holomorphic (antiholomorphic) part are given by
\begin{equation}
[J^{\alpha}_n,J^{\beta}_m]=f^{\alpha\beta}_{\gamma}J^{\gamma}_{n+m}+
\frac{k}{2}n\,g^{\alpha\beta}\delta_{n+m}\quad ,
\end{equation}
where $k$ is the level, $g^{\alpha\beta}$ is the Killing metric of
$sl_2$ and $f^{\alpha\beta}_{\gamma}$ are its structure constants.

The complete system of states (Hilbert space) involved in the theory
can be decomposed as
\begin{equation}
{\mathcal H}=\oplus_{\{j,\bar j\}}\Phi^{[j]}\otimes\Phi^{[\bar j]}\quad .
\end{equation}
Here $\Phi^{[j]}$ is a representation of $\2$.

I will only consider the diagonal embedding the Hilbert
space into a tensor product of two holomorphic spaces of states in
what follows. Such
models are known in the literature as ''A'' series. Due to this reason I will
suppress the $\bar j$-dependence as well as $\bar\Delta\,,\bar h$ etc 
below.

Let me also restrict to the case when $\Phi^{[j]}$ are the highest
weight representations of $\2$. In this case all reducible
representations are known \cite{KK}, namely, they are given by the
Kac-Kazhdan set
\begin{eqnarray}
\lefteqn{j^+_{n.m}=\frac{1-n}{2}(k+2)+\frac{m-1}{2}\quad ,}
\nonumber\\ 
\lefteqn{j^-_{n.m}=\frac{n}{2}(k+2)-\frac{m+1}{2}\quad ,}
\end{eqnarray}
with $k\in{\bf C}\,,\,\,\{n,m\}\in{\bf N}$. Note that the unitary 
representations are given by $j^+_{1.m}$ with the integer level $k$.

In general, given a representation of a symmetry algebra, to define
a field theory, one needs a construction attaching representation to a
point on a curve. In the particular case at hand, a representation
should be attached to a pair. The first parameter is a point on the
Riemann surface. As to the second, it can be taken as an isotopic
coordinate. From a mathematical point of view, this has been established
in \cite{FM}. However, this has a very simple physical
interpretation. Since $\Delta$(conformal dimension) is quadratic in
$j$(weight) one has to introduce additional parameters in order to
define OP algebra of physical fields unambiguously otherwise it is
defined up to $j=-j-1$ identification. There is no problem for the
unitary case; this is, however, not the case for a general $j$ given by
(3). 

It is surprising that the unitary models were solved by Fateev and
Zamolodchikov using this improved construction attaching representation
to a point \cite{FZ}. So it seems very natural to postulate some
basic OP expansions derived in that work as defining relations for
$SL(2)$ conformal field theories whose primary fields are
parametrized by the set (3). This was done in \cite{A1}.
 I will call such theories as the degenerate $SL(2)$ conformal 
field theories.

Define $x(\bx)$-dependent generators of $\2$ as
\begin{eqnarray}
\lefteqn{J_n^-(x)=J_n^-\,,\,\,J_n^0(x)=J_n^0+xJ^-_n\quad ,}\nonumber\\
\lefteqn{J_n^+(x)=J_n^+-2xJ^0_n-x^2J^-_n\quad .}
\end{eqnarray}
Here $x$ is an isotopic coordinate. It is easy to verify that
$J^{\alpha}_n(x)$ have the same commutation relations as
$J^{\alpha}_n$ (see (1)), i.e. they form the Kac-Moody algebra. Next
I proceed along the standard lines. Introducing the highest weight
representations $\Phi^{[j]}(x)\otimes\Phi^{[j]}(\bx)$ one
automatically generates the primary fields $\fzx{}$ together with all
their descendants. In above, $z$ is a point on the sphere. The OP
expansion for such primaries is given by
\begin{eqnarray}
\lefteqn{\fzx{1}\fzx{2}=}\nonumber\\
\lefteqn{\sum_{j_3}\frac{\vert x_{12}\vert^{2(j_1+j_2-j_3)}}
{\vert z_{12}\vert^{2(\Delta_1+\Delta_2-\Delta_3)}}
C^{j_1j_2}_{j_3}\Phi^{j_3}(x_2,\bx_2,z_2,\bz_2)\,,}
\end{eqnarray}
with $\Delta =j(j+1)/(k+2)$.
The coefficients $C^{j_1j_2}_{j_3}$ are called the structure constants
of the Operator Product algebra. It is evident that the isotopic
coordinates provides the well-defined OP algebra.

The two and three point functions of the primary fields are defined
from $SL(2)$ invariances. As to the others, they are found from the
Knizhnik-Zamolodchikov equations. General four point functions were 
derived in \cite{A1}. Moreover in this work I wrote down the structure
constants of the OP algebra (5). So the SL(2) degenerate conformal
field theories were solved.

{}From the set (3) it is worth to distinguish the so-called admissible
representations \cite{KW}, which correspond to the rational level $k$,
namely, $k+2=p/q$ with the coprime integers $p$ and $q$. In this case
there is a symmetry $j^-_{n.m}=j^+_{q-n+1,p-m}$ which allows one to
reduce the primaries parametrized by $j^-_{n.m}$ to the ones
parametrized by $j^+_{n.m}$. The OP algebra is closed in the grid
$1\leq n_i\leq q\,,\,1\leq m_i\leq p-1$. The corresponding fusion
rules are given by
\begin{eqnarray}
\lefteqn{
\vert n_{12}\vert+1\leq n_3\leq 
min\pmatrix{n_1+n_2-1\cr 2q-n_1-n_2+1}\, ,}\\
\lefteqn{\vert m_{12}\vert +1\leq m_3\leq 
min\pmatrix{m_1+m_2-1\cr 2p-m_1-m_2-1}\,,}\nonumber
\end{eqnarray}
with the following steps $\Delta n_3=1\,,\,\Delta m_3=2$.
\newline This fusion rules were first found in \cite{FZ,AY} from the
differential equations for the conformal blocks. They reveal the
quantum group structure $(U_qosp(2/1),U_qsl(2))$ of the models 
\cite{FM2}.

To complete the story on $SL(2)$, I would like to refer to recent works
\cite{SL2}.

\section{2d gravity coupled to $c\leq 1$ matter in the Polyakov
light-cone gauge}
This section attempts to briefly describe an application of the
results obtained in section two to 2d gravity (see \cite{A2} for
details).

Since the seminal works of Polyakov, Knizhnik and Zamolodchikov
\cite{KPZ}, there has been much progress in understanding the
continuum fields theory approach to 2d gravity. The majority of
efforts has been devoted to the study of coupling of conformal matter
to gravity in the conformal gauge. The reason why it is useful lies in
the fact that it is the standard gauge and its properties on the
Riemann surfaces are well known. At the same time, the properties of
the Polyakov gauge are little known which restricts the applications
of such a gauge. However it is turned out that the $SL(2)/SL(2)$
topological model reformulated in terms of the previous section
provides a way to investigate problems in the light-cone gauge. Such
model has $\2\oplus\2\oplus\2$ algebra as the symmetry algebra
\cite{GK}. The last term is a contribution of the first order
fermionic system (ghosts) of weights (1.0). The levels are given by
\begin{equation}
k_1=k\quad ,\quad k_2=-k-4\quad ,\quad k_3=4\quad .
\end{equation}
The physical fields (holomorphic part) at ghost number zero can be 
written as
\begin{equation}
\Phi^{j_1.j_2}(x,\bx,z)=\Phi^{j_1}(x,z)\Phi^{j_2}(\bx,z)\quad ,
\end{equation}
where $\Phi^j$ are the primaries of $\2$.

The idea that the $SL(2)/SL(2)$ model is connected to the minimal
models coupled to gravity was put forward in ref.\cite{Aetal}. This
discusses mainly the conformal gauge. Let me now show how it works in
my framework (with the isotopic coordinates). Setting $x=\bx=z$
and $j_2=-j_1-1$ with $j_1$ defined in (3) one immediately
obtains the minimal model coupled to gravity, more correctly only its {\it
holomorphic} sector, in the conformal gauge. It is surprising that
there exists another way, namely, by setting $x=z$ and $j_2=j_1$. As a result
one has a model ({\it holomorphic and antiholomorphic sectors}) 
which contains all features of the minimal model coupled to gravity in the
Polyakov light-cone gauge. However in contrast to the Polyakov gauge a
global structure of 2d world sheet is now well-defined that permits
one to compute correlation functions of the physical operators. The
latter are given by
\begin{equation}
{\mathcal O}_{n.m}=\int
d\mu(x,\bx;j_{n.m})\phi_{n.m}(x)\Phi^{j_{n.m}}(\bx,x)\,\, .
\end{equation}
Here $\mu(x,\bx;j_{n.m})$ represents a measure and $\phi$, $\Phi$ are
the primaries of the minimal model and SL(2) degenerate conformal
field theory, respectively. As an example, I computed the three point
functions of the physical operators ${\mathcal O}_{1.m}$ \cite{A2}. 
The results
revealed the same property as was found in the conformal gauge,
namely, the OP algebra of the physical operators is not closed anymore
\cite{GLD}.
 
\section{Some chiral rings of N=2 discrete superconformal series 
induced by SL(2) degenerate conformal field theories}

In this section I sketch a link between some $N=2$ discrete
superconformal series and $SL(2)$ degenerate conformal field theories
along the lines of ref.\cite{A3}.

The starting point is the fermionic construction proposed by Di Vecchia,
Petersen, Yu and Zheng to build the unitary representations of the
$N=2$ superconformal algebra in terms of free fermions and unitary
representations of $\2$ \cite{DPYZ}. In fact one can do better: the
only difference between the unitary representations of $\2$ and
degenerate ones is a value of $k$ (see(3)). Therefore one can relate
the degenerate representations of $\2$ to some discrete series 
of $N=2$. So it allows one to investigate a ''minimal''
non-unitary sector of the discrete series of $N=2$ (see \cite{A3} for
more details). As a result, the following relations between conformal
dimensions $h$ and U(1) charges $q$ of $N=2$ primaries in the
Neveu-Schwarz sector on the one
hand and weights $j$ and magnetic quantum numbers $\mu$ of $SL(2)$
primaries on the other hand were found
\begin{equation}
h=\frac{j(j+1)}{k+2}-\frac{\mu^2}{k+2}\quad ,\quad
q=\frac{\mu}{k+2}\quad .
\end{equation}
In the problem at hand SL(2) primaries are defined as 
\begin{eqnarray}
\lefteqn{\Phi^j_{\mu}(z,\bz)=}\nonumber\\
\lefteqn{\frac{1}{{\mathcal N}(j,\mu)}\oint_{C}\oint_{{\bar C}}
dxd\bx (x\bx)^{\mu-j-1}\Phi^j(x,\bx,z,\bz)\,\,,}
\end{eqnarray}
where $C\,,\,\bar C$ are closed contours, $\mu$ is the magnetic
quantum number and ${\mathcal N}(j,\mu)$ - normalization factors \cite{A3}.

There is also a relation between correlation functions of these
theories
\begin{equation}
\langle\prod_{i=1}^N\Phi^{h_i}_{q_i}(z_i,\bz_i)\rangle =
\prod_{i<j}^N\vert z_{ij}\vert^{\lambda_{ij}}
\langle\prod_{i=1}^N\Phi^{j_i}_{\mu_i}(z_i,\bz_i)\rangle\,, 
\end{equation}
with $\lambda_{ij}=-4\mu_i\mu_j/k+2$.

It should be stressed that the primaries fields defined in (11) depend
on contours $C({\bar C})$ in the isotopic spaces. From this point of
view one has the non-local operators. The correct contours $C_i({\bar
C}_i)$, for a particular conformal block, should be chosen by the
correct singularities at $z_{ij}\rightarrow 0$, which should match to
an OP algebra in a consistent way.

Let me restrict to the so-called primary chiral fields\footnote{This
case is easiest to analyze.}
\cite{LVW}. For such fields one has $q=h$. It simplifies integrals over
$x(\bx)$ and due to this reason one can investigate properties of the
OP algebra of the primary chiral fields \cite{A3}. It turns out that
the fields don't generate the ring. The origin of this disaster is the
non-unitarity of the models. In the case at hand the U(1) conservation
law doesn't provide a proper selection rule. It forces me to look for
more fine structures. In attempting to do this it is advantageous to
use operators introduced by Moore and Reshetikhin \cite{MR}. The point
is that a operator ${}^{\alpha}\Phi^h_q$ is associated with a triple
$(h,q,\alpha)$, where $h$ and $q$ are the conformal dimension and U(1)
charge. As to $\alpha$, it means a pair of states in the highest weight
representations of the quantum groups $(U_qosp(2/1),U_qsl(2))$. If the
states $\alpha$ are the highest weight vectors then the operators
${}^{\alpha}\Phi^h_h$ define the ring \cite{A3}. This solution
provides a strong evidence that a quantum group underlies the ring. It
is disguised in the unitary case in virtue of the U(1) conservation
law, but it becomes clear in the non-unitary case. 

\section{Conclusions and remarks}

First, let me say a few words about results.

In the above I have briefly outlined the $SL(2)$ degenerate conformal
field theories and their applications to the 2d gravity in the
Polyakov light-cone gauge and some $N=2$ discrete superconformal
series. The main moral of the story is the isotopic coordinates
$x(\bx)$. On the one hand they provide the well-defined OP algebra and
enlarge the degree of applications. On the other hand, a natural
question arises: is the theory really two-dimensional or it is a
restriction of a certain four dimensional one? Unfortunately at this moment
I don't know of an exact answer to this magical question.

Let me conclude by mentioning some open problems.
\newline$\bullet$ An important problem which wasn't discussed 
in \cite{A1} is to
check that the solutions of the KZ equations also satisfy a system of
equations which follows from the singular vectors in the highest
weight representations of $\2$ parametrized by the Kac-Kazhdan set
(3).
\newline$\bullet$ The next open problem is to solve non-diagonal 
$SL(2)$ theories.
\newline$\bullet$ Due to the solution of the $SL(2)$ degenerate conformal field
theories, there is a strong indication on a finite number of order
parameters in a ''parafermionic'' theory for a rational $k$. The
problem is to investigate such coset $SL(2)/U(1)$ in more
detail. Furthermore there exists another problem, namely, to find
models of statistical mechanics which have fixed points described by the
coset $SL(2)/U(1)$.
\newline$\bullet$ The main problem in the context of 2d gravity 
is, of course, to compute four point correlation functions.
\newline$\bullet$ As to the $N=2$ discrete superconformal theories they are
waiting to be solved.

{\bf{Acknowledgments}}
It is a pleasure to thank B.Feigin, J.-L.Gervais, R.Metsaev and
A.Schwarz for useful discussions and G.Lopes Cardoso for reading the
manuscript, and 
of course D.L\"ust, H.-J.Otto and
G.Weigt for organizing Buckow Symposium 96. The hospitality extended to
me at Laboratoire de Physique Th\'eorique de l'\'Ecole Normale 
Sup\'erieure, where these notes were written, is acknowledged. 
This research was supported
in part by Landau-ENS Jumelage, INTAS grant 94-4720, and by 
Russian Basic Research Foundation under grant 96-02-16507.


\begin{thebibliography}{99}

\bibitem{BPZ} 
A.A.Belavin, A.M.Polyakov and A.B. Zamo- lodchikov, 
Nucl.Phys. B241 1984 333.

\bibitem{DF} 
Vl.S.Dotsenko and V.A.Fateev, Nucl.Phys. B240 [FS12] (1984) 312;
Nucl.Phys. B251 [FS13] (1985) 691; \PL{157}{1985} 291.

\bibitem{FZ} 
V.A.Fateev and A.B.Zamolodchikov, Sov.J. Nucl.Phys. 43 (1986) 657.

\bibitem{KK}
V.G.Kac and D.A.Kazhdan, Adv.Math. 34 (1979) 97.

\bibitem{FM}
B.Feigin and F.Malikov, Lett.Math.Phys. 31 (1994) 315.

\bibitem{A1}
O.Andreev, \PL{363}{1995} 166.

\bibitem{KW}
V.G.Kac and M.Wakimoto, Proc.Natl.Acad. Sci.USA 85 (1988) 4956.

\bibitem{AY}
H.Awata and Y.Yamada, Mod.Phys.Lett. A7 (1992) 1185.

\bibitem{FM2}
F.Malikov and B.Feigin, Modular Functor and Representation Theory of
$\2$ at a Rational Level, q-alg/9511011.

\bibitem{SL2}
J.L.Petersen, J.Rasmussen and M.Yu, Monodromy invariant Green's
functions in WZNW theories with fractional level, 
Preprint AS-ITP-96-26; Fusion, crossing and monodromy in
conformal field theory based on SL(2) current algebra with fractional
level, hep-th/9607129; \\
J.Rasmussen, Applications of free fields in 2D current algebra,
hep-th/9610167;\\
P.Furlan, A.Ch.Ganchev and V.B.Petkova, $A_1^{(1)}$ admissible
representations: fusion transformations and local correlators,
hep-th/9608018.

\bibitem{A2}
O.Andreev, \PL{375}{1996} 60.

\bibitem{KPZ}
A.Polyakov, Mod.Phys.Lett. A2 1987 893;
in Les Houches 1988: Two-dimensional quantum
gravity. Superconductivity at high $T_c$; \\
V.Knizhnik, A.Polyakov and A.Zamolodchi- kov, Mod.Phys.Lett. A3 1988 819.

\bibitem{GK}
K.Gawedzki and A.Kupianen, Phys.Lett. B215 (1988) 119, Nucl.Phys. 
B320 (1989) 649.

\bibitem{Aetal}
O.Aharony, O.Ganor, J.Sonnenschein, S.Yan- kielowicz and 
N.Sochen, Nucl.Phys. B399 (1993) 527; \\
H.Hu and M.Yu, Nucl.Phys.B391 (1993) 389.

\bibitem{GLD}
M.Goulian and M.Li,Phys.Rev.Lett.66 (1991) 2051;\\
Vl.S.Dotsenko, Mod.Phys.Lett.A6 1991 3601.

\bibitem{A3}
O.Andreev, Some chiral rings of N=2 discrete superconformal series 
induced by SL(2) degenerate conformal field theories, hep-th/9612043.

\bibitem{DPYZ}
P.Di Vecchia, J.L.Petersen, M.Yu and H.Zheng, \PL{174}{1986} 280.

\bibitem{LVW}
W.Lerche, C.Vafa and N.Warner, \NP{324}{1989} 427.

\bibitem{MR}
G.Moore and N.Reshetikhin,Nucl.Phys.B328 (1989) 557.

\end{thebibliography}
\end{document}